\documentclass[aip,amsmath,amssymb,reprint,sort&compress]{revtex4-1}
\usepackage{hyperref}

\usepackage{graphicx} 
\usepackage{dcolumn} 
\usepackage{bm} 
\usepackage[utf8]{inputenc}
\usepackage[T1]{fontenc}
\usepackage{mathptmx}
\usepackage{xcolor} 
\usepackage{braket} 

\begin{document}

\title{Dissipative Forces in Photon-Medium Interactions\\Using Perturbation Theory}

\author{Raju S. Khatiwada}
\affiliation{GoldenGate International College, Tribhuvan University, Kathmandu, Nepal}
\email{r.khatiwada@goldengate.edu.np}

\author{N. P. Adhikari}
\affiliation{Central Department of Physics, Tribhuvan University, Kirtipur, Kathmandu, Nepal}

\author{C. Ortiz}
\affiliation{Unidad Acad\'emica de F\'isica, Universidad Aut\'onoma de Zacatecas, 98060, Zacatecas M\'exico}

\date{27 November 2024}

\begin{abstract} 

This study examines dissipative forces in photon-medium interactions through time-independent perturbation theory, with a specific focus on single Helium-4 atoms. Utilizing a Hamiltonian framework, energy corrections induced by dissipative gravitational frictional effects in low-density systems are derived and analyzed as a function of inter-atomic distance. The calculations reveal an energy correction peak at $r_1 = 0.1 nm$, followed by rapid exponential decay, highlighting the dominance of nonlinear dissipative effects at nanoscale separations. These findings emphasize the critical role of short-range interactions, governed by the de-Broglie wavelength of Helium-4, and provide a rigorous theoretical basis for understanding photon-medium interactions at quantum scales. This novel single-particle approach departs from macroscopic mean-field models, offering unique insights into the microscopic mechanisms underlying energy dissipation. The results have potential implications for advancing quantum information processing, nonlinear optics, and the study of dissipative mechanisms in quantum fluids. Experimental validation of the theoretical predictions is proposed using state-of-the-art techniques in optical media, levitated nanoparticle systems, and integrated photonic circuits.

\end{abstract}

\maketitle

\section{Introduction} \label{sec1}
Helium-4 is a unique fluid that exhibits quantum properties at low temperatures, making it a widely studied system in both theoretical and experimental physics. At temperatures below 2.17 K, Helium-4 undergoes a phase transition to a superfluid state, characterized by zero viscosity and the ability to flow without resistance. This superfluidity, along with the formation of quantized vortices, makes Helium-4 an ideal system for studying nonlinear and dissipative forces in quantum fluids. The work of researchers like Donnelly (1991) \cite{Donnelli} has extensively documented these collective behaviors, focusing on macroscopic properties such as vortex dynamics and the complex interplay of quantum effects. Feynman’s 1955 study also provided foundational insights by suggesting that turbulence in superfluid Helium is associated with collisions between thermal excitations and quantized vortices \cite{FEYNMAN195517}.

Despite these advancements, much of the existing research on Helium-4 relies on collective models rather than individual atomic interactions. Models such as the Gross-Pitaevskii equation advance our understanding of Bose-Einstein condensates and superfluid phenomena, but they are designed to capture many-body effects at a macroscopic scale\cite{1961NCim...20..454G,Pitaevskii_1961}. These models emphasize mean-field approximations, which average out individual particle interactions and primarily focus on bulk properties. Consequently, while these collective approaches have been highly successful, they offer limited insight into the behavior of individual Helium atoms and their interactions with other particles and photons.

Studying single-particle dynamics in Helium-4 allows us to analyze dissipative forces at a finer level, revealing mechanisms of energy loss and momentum transfer that are often obscured in bulk models. This single-particle approach provides a complementary perspective, enhancing our understanding of energy dissipation within atomic systems by focusing on the isolated interactions.

The resolution of momentum transfer debates, such as the Abraham-Minkowski controversy \cite{abraham1959momentum,minkowski1908beitrag}, is particularly relevant at atomic scales, where the interplay of radiation pressure and dissipative forces governs particle dynamics. It is a well-established fact that light exerts radiation pressure when it interacts with a medium\cite{jones1978radiation}. This momentum transfer becomes particularly relevant when studying light-matter interactions at atomic scales, where even minor shifts in photon energy or momentum could affect particle dynamics. Although radiation pressure has been thought to be small enough, it was only after the invention of the laser in 1960 that this field garnered significant interest.  Further exploration of this phenomenon involving momentum changes of the photon has contributed to Optical trapping mechanisms \cite{PhysRevLett.24.156}.

Classical wave equations, commonly used to describe wave propagation in physical media, are ineffective when applied to photon interactions in quantum fluids. These equations typically do not account for nonlinear dynamics and quantum-specific effects critical to describing energy dissipation at atomic scales. Traditional wave equations, like those used in acoustics or electromagnetics, assume linear and often static media, whereas the interaction of photons with quantum fluids like Helium-4 involves complex, time-dependent, and nonlinear phenomena. By adopting a Hamiltonian framework \cite{ZAKHAROV1985285} and applying perturbation theory, we can model this photon-medium interaction as weak perturbations to a known quantum state. In this study, we employ the ground-state wavefunction of a single Helium-4 atom to examine energy corrections associated with dissipative forces. Based on non-degenerate perturbation theory, this approach allows us to quantify how energy dissipation may occur at atomic scales.

In this study, the “low-density” medium implies that particle spacing in the medium is sufficiently large to minimize inter-particle interactions, allowing us to approximate each particle’s behavior independently. We model photon-medium interactions as weak perturbations to a known quantum state. This approach offers new insights into the mechanisms of energy dissipation at atomic scales, bridging a gap between collective models and single-particle dynamics. The findings contribute to a deeper understanding of nonlinear and quantum-specific effects in light-matter interactions, with potential applications in quantum optics and condensed matter physics.

The paper is structured as follows: In Section 2, we begin by examining the foundational physics governing photon interactions in low-density media. Section 3 develops the mathematical framework, deriving first-order energy corrections using a perturbative Hamiltonian and Helium-4's ground-state wavefunction. Section 4 presents analytical results, emphasizing interactions tied to the de Broglie wavelength, while Section 5 proposes possible experimental validation techniques. Finally, in Section 6, we discuss future research directions and challenges isolating dissipative forces at quantum scales.

\section{Photon Interactions in Low-Density Medium} \label{sec2}

The interaction of photons in low-density media involves nonlinear and dissipative effects that alter photon dynamics without relying on large-scale, collective behavior. Such photon-atom interactions are amplified in systems where strong light-matter coupling is present. For example, in studies of polaritons within cavity quantum electrodynamics (QED), Carusotto and Ciuti \cite{Carusotto_2013} demonstrated that photon-phonon interactions in polaritonic systems can lead to nonlinear optical effects that modulate the medium’s optical properties without requiring explicit scattering. Similarly, optomechanical systems have shown that photon-phonon interactions can modify both optical and mechanical properties, as observed by Primo et al. (2023) \cite{Primo2023}.

To contextualize the short-range energy corrections in our study, it is instructive to examine the Uehling potential, a quantum mechanical correction to the Coulomb interaction between charged particles arising from vacuum polarization effects in quantum electrodynamics \cite{1935PhRv...48...55U}. This correction becomes prominent at distances comparable to or smaller than the electron's Compton wavelength ($\lambda_c \approx 3.86 fm$), as it originates from the polarization of the vacuum due to virtual electron-positron pairs \cite{Frolov2024}. The Uehling potential demonstrates that minute quantum effects, often negligible at larger scales, can create observable shifts in particle interactions at short distances. These shifts provide insights into the fundamental behavior of forces within atomic systems across other different distances.

The primary mechanism behind photon-medium interactions in Helium-4 atoms involves the coupling of single photons to single atoms, typically facilitated by advanced optical techniques. Recent studies have explored the use of super-resolution imaging techniques, such as 4Pi microscopy \cite{jen2024photonmediateddipoledipoleinteractionsresource}, which surpass the diffraction limit, enabling more effective light-atom coupling. In these experiments, significant levels of extinction of the incident light field have been observed, indicating a strong nonlinear interaction at the single-photon level. Photon-mediated dipole-dipole interactions (PMDDIs) also play a crucial role in these processes.

From R. J. Donnelly’s work on Quantized Vortices \cite{Donnelli}, low-energy excitations such as phonons and higher-energy excitations like rotons can be treated as independent particles or quasi-particles that follow their own dynamics within the superfluid. At absolute zero, the quasi-particle model pictures the superfluid as forming a continuous fluid, and the single wavefunction ($\psi_k $) describes the fluid\cite{Donnelli}. Further, Brooks \cite{Brooks}, in 1973 with his dispersion curves for Helium atoms, observed that the phonons dominate the density fluctuations at longer wavelengths. This justifies the use of a single wavefunction to represent the Helium-4 atom.

Since the perturbations are weak and we are only interested in finding the energy corrections, we use time-independent perturbation theory. The entire field, represented by the wavefunction $\psi(r,t)$, is subject to the weak perturbation of the nonlinear medium as represented by equation (\ref{H}). Also, the system is non-degenerate since the phonon wavevector $k$ leads to a distinct energy eigenvalue.

When the perturbations are strong and time-varying, the time-dependent perturbation theory should be used.
The work involving time evolution of the wavefunction $\psi(r,t)$ due to nonlinear forces is in progress and is similar to the one by Polkovnikov \cite{Polkovnikov}. It will be published elsewhere in the future.

\section{Mathematical Framework} \label{sec3}

To model the behavior of photons in a quantum fluid, we begin by approximating the unperturbed ground state wavefunction of the Helium-4 atom.

\subsection{Ground State Wavefunction}

The phonon wavefunction of the Helium 4 atom ($\psi_k $) represents a collective excitation within a system of particles, where $\psi_0$ could be seen as the ground state wavefunction of each atom in the medium.

\begin {equation}
 \psi_k =Const. \sum^N_{j=1}e^{ik.r_j}\psi_0.
 \end{equation}

\begin{figure}[h]
    \centering
    \includegraphics[width=0.4\textwidth]{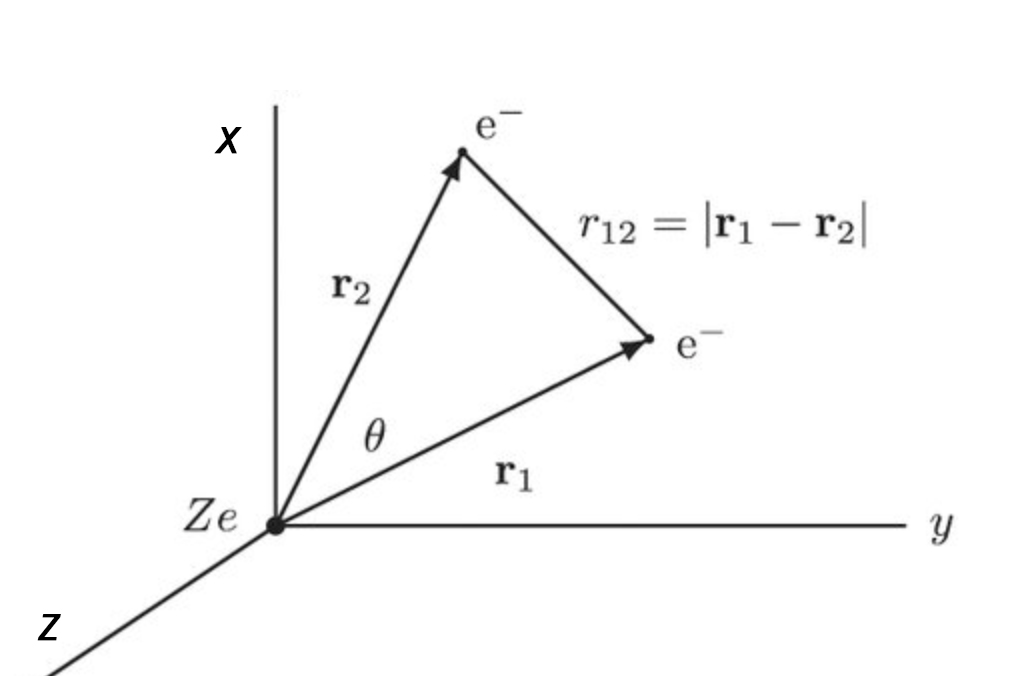} 
    \caption{The Helium Atom.}
    \label{fig:example} 
\end{figure}

The wavefunction $\psi_k$ is a superposition of single-particle ground states, each shifted by a phase factor corresponding to the phonon wavevector $k$. By transitioning to the single-atom ground state wavefunction, we can avoid the complexities associated with collective behavior, simplifying the calculations to focus solely on perturbations affecting a single Helium atom and focus on our study on energy corrections and dissipative forces at atomic scales.

In liquid Helium, the ground state wavefunction isn't constant but achieves a maximum amplitude when the particles are uniformly spread apart because of hard-core repulsion between the atoms.

If we ignore the electron-electron repulsion, the Hamiltonian splits into that of two Hydrogen atoms with only the nuclear charge replaced to $2e$ instead of $e$.

In that case, the actual wavefunction is just the product of two hydrogen wavefunctions.
 \begin{equation}
 \psi_0(r_1, r_2)= \psi_{1 0 0}(r_1)\psi_{{1} {0} {0}}(r_2) = \frac{8}{\pi a^3}e^{\frac{-2(r_1+ r_2)}{a}}
\end{equation}

Since we also want to account for electronic repulsion between electrons in the helium atom, we cannot take the wave function above. The effective nuclear charge has to be taken into account as well. For some effective value of Z, the wavefunction that corresponds to the experimental value of the ground state energy of $-79$ eV is given by:

 \begin{equation}
 \psi_0(r_1, r_2)=  \frac{Z^3}{\pi a^3}e^{\frac{-Z(r_1+ r_2)}{a}}.
\end{equation}
The electronic repulsion shields the nuclear charge. So, naturally, Z can be expected to be less than '2'. In the interface of Helium-4 atoms, there is a weak inter-atomic interaction when forces between all constituent particles are considered. This gives rise to the London dispersion force. Further there is also a quantum mechanical attractive force due to the overlapping of the orbitals. At its core, this is due to the symmetric nature of the wavefunction of the bosons. According to Griffiths, $Z=1.69$ from the variational principle when $H$ is minimized\cite{Griffiths2004}.

\subsection{Perturbing Hamiltonian} \label{perturbativehamiltonian}

In this paper, Ortiz et al.\cite{Ortiz2023} demonstrated how applying d’Alembert’s principle of virtual work provides a formal method to establish energy dissipation due to gravitational fields in a low-density medium when non-holonomic constraints are present. Usually, dissipativity is limited to low-density media where the mean free path of the particle exceeds its path displacement. In denser media, energy dissipation is dominated by conventional loss mechanisms, such as scattering and absorption, which obscure the subtle energy losses due to dissipative processes. This, along with the constant-velocity constraint of photons in a medium, can be overcome using d’Alembert’s principle, which incorporates virtual work to address non-holonomic constraints. We can employ a similar approach for photons traveling through a quantum fluid like Helium-4. The force on a particle due to the medium is given as:

\begin{equation}
    F= -\frac{2}{3}\pi G m_0 \rho v_r t.
\end{equation}

This force shapes the perturbing Hamiltonian because it accounts for the nonlinear interaction between the photon and the medium. With this perturbing Hamiltonian defined, we can explore the non-linear realm of photon-medium interaction on an atomic scale.

In this case, placing the constraint equation $r=v_r t$ and integrating from 0 to r, the Hamiltonian that causes the perturbation is:

\begin{equation}\label{H}
    H^{'} = -\frac{\pi }{3} Gm_0 \rho_0 r^2.
\end{equation}

Here, $\rho_0$ represents the density, assumed constant due to the homogeneity of the medium.

The nonlinearity arises from the quadratic dependence on the photon's position. In this paper by Suassuna et al.\cite{Suassuna}, a nonlinear perturbative correction could be represented as the product of a feedback gain $G_{fb}$ and a nonlinear function of the particle's position, $f(r)$. This further supports our treatment of nonlinear perturbation, which involves a quadratic dependence on distance.

Moreover, studies on levitated nanoparticles by Kremer et al. \cite{Kremer} measured shifts in the power spectrum of particle motion due to nonlinear forces acting as perturbations. These experimental findings highlight the importance of nonlinear dynamics in optical trapping systems. We discuss more experimental scope in section \ref{sec5}.

Thus, our formulation provides a solid foundation for exploring perturbative effects in photon-medium interactions for any perturbing Hamiltonian.

\subsection{Total Hamiltonian and Perturbation Theory}

The total Hamiltonian acting on the photon is the sum of the kinetic, potential, and perturbative terms:
\begin{equation}
    H = \frac{1}{2} m_0 \dot{r}^2 - \frac{G m_0 M}{r} - \frac{\pi}{3} G m_0 \rho_0 r^2.
\end{equation}

Using non-degenerate perturbation theory, the first-order energy correction is given by the expectation value of the perturbing Hamiltonian in the ground state:
\begin{equation}
    E'_n = \langle \psi^0_n | H' | \psi^0_n \rangle.
\end{equation}
Substituting the perturbing Hamiltonian from Eq. [\ref{H}], the energy correction becomes:
\begin{equation}\label{correction}
    E' = \left(\frac{Z^3}{\pi a^3}\right)^2 \frac{\pi}{3} G m_0 \rho_0 \int e^{-\frac{2Z(r_1 + r_2)}{a}} |\mathbf{r}_1 - \mathbf{r}_2|^2 \, d^3r_1 \, d^3r_2.
\end{equation}

\subsection{Integration in Spherical Coordinates}

We use spherical coordinates to simplify the integrals, as they involve the modulus of vectors $r_1$ and $r_2$ and will parameterize both vectors accordingly. We fix $r_1$ so that the polar axis lies along $r_1$.

For simplicity, the above equation (\ref{correction}) can be written as:

\begin{equation} \label{E9}
    {E}'= A \int I_2 e^{-\frac{2Zr_1}{a}} d^3r_1,
\end{equation}
where $A = (\frac{Z^3}{\pi a^3})^2 \frac{\pi }{3} Gm_0 \rho_0$ and $I_2$, defined as integral over $r_2$, is given as:
\begin{equation}
    I_2=\int e^{-\frac{2Zr_2}{a}} \left\vert r_1-r_2\right\vert^2  d^3r_2.
\end{equation}

By the law of cosines, $\left\vert r_1-r_2\right\vert = \sqrt{r_1^2 +r_2^2 - 2r_1r_2cos\theta_2}$ and $d^3r_2 = r_2^2sin\theta_2dr_2d\theta_2d\phi_2$. Therefore $I_2$ becomes:

\begin{equation}
       I_2= \int e^{-\frac{2 Z r_2}{a}} ({r_1^2 +r_2^2 - 2r_1r_2cos\theta_2})r_2^2sin\theta_2dr_2d\theta_2d\phi_2.
\end{equation}

The $\phi_2$ integral is trivial $(2\pi)$ and the $\theta_2$ integral is given as:

\[
\int_0^\pi \sin\theta_2({r_1^2 +r_2^2 - 2r_1r_2cos\theta_2})d\theta_2 = 2(r_1^2 + r_2^2).\]

Thus, using the $\phi_2$ and $\theta_2$ integrals,
\begin{equation} \label{I2final}
    I_2 = 4 \pi r_1^{2}\int_0^{r_1}[1+(\frac{r_2}{r_1})^2] e^{-\frac{2Zr_2}{a}}r_2^2dr_2.
\end{equation}

After substituting A and $I_2$ in Eq. (\ref{E9}), we get an energy correction expression.

\begin{align}
   E' = & \;  \left( \frac{Z^3}{\pi a^3} \right)^2 \frac{\pi}{3} G (\frac{h}{\lambda c}) \rho_0 (4\pi)^2 \int_0^{\infty} {r_1}^4 e^{-\frac{2Zr_1}{a}} \notag \\
   & \times \int_0^{r_1} \left[1+\left(\frac{r_2}{r_1}\right)^2\right] e^{-\frac{2Zr_2}{a}} {r_2}^2 \, dr_2 \, dr_1.
\end{align}

Notice that we are replacing $m_0$ by the relation involving the associated momentum and the de Broglie wavelength as:

\begin{equation}
    m_0=\frac{h}{\lambda c}
\end{equation}

Since wavelength $\lambda$ and momentum $p$ are affected by the refractive index and dispersion of the medium, the revised equation can naturally incorporate these effects.

Finally, after integration and applying numerical techniques where $r_1 >0$, we get the energy correction per unit momentum.

\begin{equation}
    E' \approx (\frac{Z^3}{\pi a^3})^2 \frac{\pi }{3} G (\frac{h}{\lambda c}) \rho_0(4\pi)^2\frac{15 _1 ^{8}}{32Z^{8}}.
\end{equation}

Because of the exponential terms in the equation, it is necessary to emphasize the short-range interactions. A finite cut-off distance for $r_1$ can be introduced. This observation is consistent with the fact that, near absolute zero, the de Broglie wavelength for lighter isotopes like Helium-II exceeds the mean free path. This is also the reason why Helium-4 is referred to as quantum fluid among many other isotopes of Helium. Thus, analytically, we set $r_1$, the effective distance, equal to the de Broglie wavelength of the Helium-4, which is denoted by $\lambda$.

Thus, the final equation becomes:

\begin{equation}
    E' \approx (\frac{Z^3}{\pi a^3})^2 \frac{\pi }{3} G (\frac{h}{\lambda c}) \rho_0(4\pi)^2\frac{15 \lambda ^{8}}{32Z^{8}}.
\end{equation}

\begin{table}[h]
    \centering
    \caption{Key Parameters and Their Values.}
    \label{tab:parameters}
    \begin{tabular}{|c|c|}
        \hline
        Parameter & Value \\
        \hline
        Effective Nuclear Charge, $Z$ & 1.69 \\
        Bohr Radius, $a$ & $5.29177 \times 10^{-11}$ m \\
        Gravitational Constant, $G$ & $6.67430 \times 10^{-11}$ m$^3$/kg/s$^2$ \\
        Medium Density, $\rho_0$ & $145$ kg/m$^3$ \\
        de Broglie wavelength, $\lambda$ & $ 0.5$ nm \\
        \hline
    \end{tabular}
\end{table}
 It is important to stress on the values of density, $\rho$, and the de Broglie wavelength of helium atoms in liquid helium, $\lambda$. Assuming the homogeneity of the medium, the density of Helium-4 is roughly $145$  $kg/m^3$, and the de Broglie wavelength is in $nm$ scale. The de Broglie wavelength can provide a quantum mechanical length scale that describes the wave-like nature of the Helium-4 atoms. Thus:

\begin{equation}
E' \approx  10^{-53} \, \text{J} \approx  10^{-34} \text{eV},
\end{equation}
is the approximate calculated value of the energy correction, using the values of the parameters from the table \ref{tab:parameters}.

This value, though very small, aligns with expectations as it represents the energy correction arising from the dissipative gravitational frictional effect. Notably, the correction depends strongly on the parameter $\lambda$ with $E'$ scaling as $\lambda^{7}$. For instance, reducing $\lambda$ from 
0.5 $nm$ (de-Broglie wavelength for Helium-4) to shorter values, while theoretically feasible, may face practical challenges due to absorption and dispersion in the medium. Nevertheless, this dependence highlights the theoretical importance of $\lambda$ in determining the strength of dissipative effects, with smaller $\lambda$ potentially enhancing the correction under idealized conditions.

It is important to stress that this GF-induced mechanism fundamentally differs from other energy shift phenomena like the Lamb shift and the Uehling potential. So, the earlier comparison is only for the sake of quantitative assessment.

\section{Discussions} \label{sec4}

\begin{figure}[ht]
\centering
\includegraphics[width=0.5\textwidth]{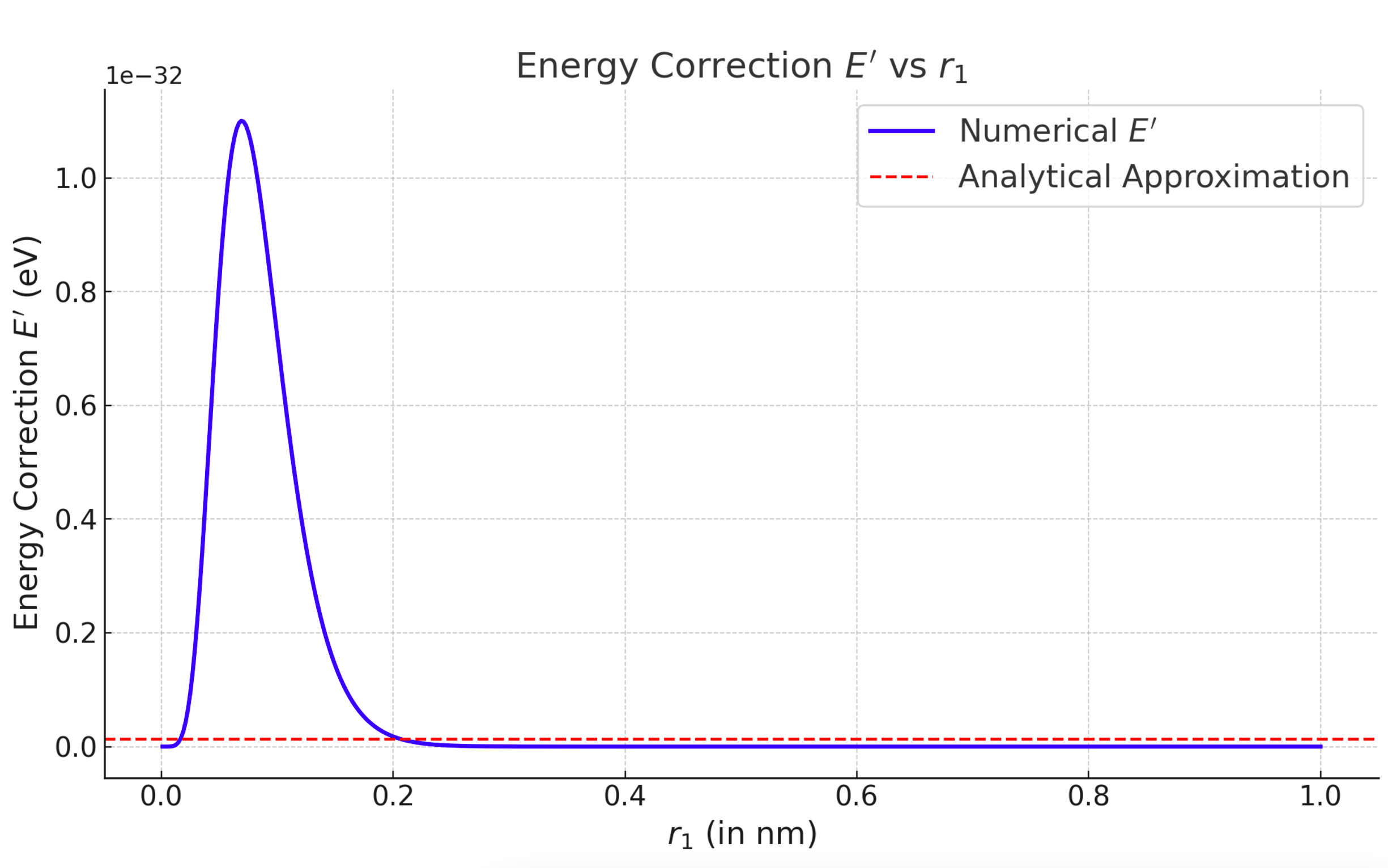}
\caption{Perturbative Dissipation V/s Distance}
\label{fig}
\end{figure}

The dominance of short-range interactions is a key feature of the derived energy correction, as evident from the exponential terms in the perturbative Hamiltonian. These interactions sharply decay with increasing distance, as shown in Fig. \ref{fig}. As $r_1$ increases beyond approximately $0.1  
nm$, the energy correction diminishes rapidly, highlighting that the mechanism is relevant only at nanometer scales or smaller. This observation aligns with the quantum nature of Helium-4, where the de-Broglie wavelength exceeds the mean free path near absolute zero. Thus, the analytical approximation assumes that the dominant contribution to the integral comes from short-range interactions ($r_1 = \lambda$), denoted by the horizontal line in the graph.

The parameter $r_1$ can also be understood as a characteristic length scale that governs the interaction range between photons and phonons, likely linked to the phonon correlation length or the mean inter-phonon distance. The inequality 
$r_1 k\ge 1$, when expressed in terms of wavenumber $k$, reflects the quantization of momentum in the medium. This condition ensures that photon wavelengths are short enough to allow efficient energy transfer.

The calculated energy correction aligns with the precision required to explore weak perturbative forces in quantum systems. Although the magnitude of $E'$ appears negligible in practical terms, it demonstrates the subtle role of dissipative forces at nanoscopic length scales. This mechanism becomes particularly relevant in localized interactions, where the dominance of short-range effects is more pronounced. Unlike collective models that average individual interactions, the single-particle framework offers a finer view of energy dissipation. Exploring cumulative effects could further amplify the correction, as discussed in the Future Scope and Challenges section.

Furthermore, this perturbative framework can be adapted to other systems where similar dissipative interactions may play a role. Extending this methodology could enable future studies to examine dissipative processes in various contexts, offering a deeper understanding of complex photon-medium interactions.

\section{Experimental Outlook} \label{sec5}
While the energy correction derived is small, its dependence on parameters such as wavelength, density, and medium properties highlights potential scalability under modified conditions, paving the way for further theoretical studies and experiments.

This section explores three key approaches: resonance-based systems, such as photonic crystals and optical cavities, which enhance sensitivity to small energy shifts; nonlinear optical techniques, including soliton dynamics, that magnify dissipative effects; and cryogenic systems, like superfluid helium, which provide ideal conditions to observe enhanced quantum effects. Each approach is explained in brief in the following subsections.

\subsection{Nonlinear Optical Media}

Nonlinear optical media, such as lithium niobate or photonic crystals, can form solitons \cite{davydov1985solitons}. Our model's predictions can be tested by inducing perturbations in solitons using phase modulation techniques. By inducing phase-modulated perturbations in solitons, one can replicate the dissipative forces described in our theoretical framework.

Experiments could use femtosecond lasers to generate ultrafast pulses to track soliton stability and energy loss. The experimental results could validate our predictions on energy dissipation rates by comparing dissipation at different photon wavelengths. The experimental findings of Bao et al. \cite{Bao2015-me} on dissipative solitons, where a controlled radio signal is applied as a perturbation in a low-density medium, provide a precedent for testing similar models. Similar techniques could be adapted to validate the dissipative forces acting on photons interacting with single helium atoms.

\subsection{Levitated Nanoparticle Systems}

Building on the work of Kremer et al. \cite{Kremer}, levitated nanoparticles in optical traps remain a versatile platform for studying nonlinear and dissipative forces, even at the scale of single helium atom interactions. In this context, the perturbative dissipative forces acting on photons can be simulated by applying controlled optical feedback to manipulate the nanoparticles. This allows us to mimic the Hamiltonian derived for single-atom photon-medium interactions.

The motion and energy dissipation of levitated nanoparticles can be precisely monitored to test the validity of our perturbative framework. By systematically varying medium density, temperature, or the optical feedback mechanism, the experimental setup can emulate the conditions under which single helium atoms interact with photons. These experiments offer a practical pathway to empirically validate the proposed energy corrections and their dependence on nanoscale parameters.

Such controlled environments enable isolation of the photon-medium interaction, ensuring that such nonlinear dissipative effects can be probed with high sensitivity. This approach bridges the gap between theoretical predictions and experiments, reinforcing the applicability of levitated nanoparticle systems to validate energy dissipation at the single-atom level.

\subsection{Integrated Photonic Circuits}
Integrated photonic circuits remain a powerful and precise platform for investigating photon interactions in confined geometries, making them well-suited for testing our predictions involving single helium atoms. These circuits enable the study of short-range photon-medium interactions by confining photons in optical resonators, where resonance shifts and Q-factor degradation can provide direct evidence of dissipative forces at atomic scales \cite{haroche2006exploring}.

The interactions described in our model can be effectively simulated within photonic circuits by fine-tuning the resonator geometry and material properties to enhance sensitivity to perturbative effects. These adjustments are particularly critical for probing the energy dissipation mechanisms predicted in the context of single helium atom interactions.

Superconducting photonic circuits operating at cryogenic temperatures are especially relevant, as they closely mimic the physical conditions of low-temperature helium systems. By leveraging their ability to confine photons in well-defined modes, these circuits can detect minute shifts in energy dissipation, validating our theoretical framework for nonlinear dissipative interactions.

\section{Future Scope and Challenges} \label{sec6}

Future theoretical work could focus on extending the framework to include cumulative or collective effects, which may amplify the impact of the mechanism in denser media or under stronger perturbative regimes. Additionally, exploring time-dependent perturbation theory could provide insights into the evolution of dissipative forces in dynamic systems.

Further research directions include experimental validation of these findings, as well as deeper insights into energy transfer mechanisms in quantum systems. Progress in these areas may lead to advances in photonic devices and contribute to a more refined understanding of nonlinear optical media, including the development of efficient quantum communication systems and advanced quantum computing architectures. However, challenges remain in isolating the effects of dissipative forces within complex atomic systems and achieving the precision necessary to detect subtle energy corrections at small scales.

Applying time-independent perturbation theory to Helium-4 in the context of photon interactions presents several specific challenges. Firstly, Helium-4 is a quantum many-body system, which inherently involves complex interactions among its particles. The perturbation theory typically assumes that the perturbation is weak compared to the unperturbed system. However, the photon-medium interactions in Helium-4 can introduce significant perturbations, complicating the calculations and potentially requiring higher-order corrections to achieve accurate results. Secondly, the photon interactions with Helium-4 atoms can involve processes such as absorption, emission, and scattering, each of which can affect the system under consideration. Accurately modeling these interactions requires a detailed understanding of the Helium-4 energy levels and transition rates, which may not be straightforward to incorporate into a time-independent perturbation framework.

Experiments employing advanced techniques such as 4Pi microscopy have demonstrated effective coupling between single photons and single atoms, leading to observable nonlinear interactions at the single-photon level. This advancement is crucial for the control and design of these experiments\cite{jen2024photonmediateddipoledipoleinteractionsresource}.

\section{Conclusions} \label{sec7}

This study establishes a theoretical framework for analyzing dissipative gravitational friction effects in photon-medium interactions, emphasizing single-particle dynamics within Helium-4 systems. The derived energy corrections reveal the underlying nonlinear mechanisms governing photon behavior at atomic scales. By employing a Hamiltonian approach combined with time-independent perturbation theory, the study emphasizes the critical role of short-range interactions and their dependence on the de-Broglie wavelength of Helium-4. These results highlight the complex relationship between photon momentum and medium characteristics, providing a novel perspective on photon-mediated energy dissipation and its sensitivity to key parameters such as density and wavelength.

The findings open avenues for exploring dissipative phenomena further, particularly through refinements that include collective effects or applications in denser media, where these mechanisms might be amplified. Experimental methods such as optical trapping, photonic circuit integration, or cryogenic systems offer promising pathways to validate the theoretical predictions and enhance their applicability. Such efforts not only have the potential to refine our understanding of photon-medium interactions but also pave the way for advancements in quantum optics, nonlinear dynamics, and emerging photonic technologies.

\section{Acknowledgements}
The authors express their sincere gratitude to Prof. Homnath Poudel of GoldenGate International College, Tribhuvan University, for his invaluable feedback on this project and for his generous assistance in facilitating access to essential resources.

\section*{References}
\bibliographystyle{unsrt}
\bibliography{reference}

\begin{widetext}
\newpage
\section*{Appendix}
The $I_2$ term after integration yields 

\begin{equation}
    I_2 = \left(\frac{a^{3} z^{2} {r_1}^{2} + 3a^{4} z{r_1} + 3a^{5}}{4z^{5} {r_1}^{2}} - \frac{\left(4az^{4} {r_1}^{4} + 6a^{2} z^{3} {r_1}^{3} + 7a^{3} z^{2} {r_1}^{2} + 6a^{4} z{r_1} + 3a^{5}\right) \mathrm{e}^{-\frac{2z{r_1}}{a}}}{4z^{5} {r_1}^{2}}\right)
\end{equation}

The energy term in the final form of integration looks like:

\begin{equation}
\begin{aligned}
     E' = (\frac{Z^3}{\pi a^3})^2 \frac{\pi }{3} G \frac{h}{\lambda c} \rho_0(4\pi)^2\int_0^\infty {r_1}^{4} \mathrm{e}^{-\frac{2z{r_1}}{a}} \left(\frac{a^{3} z^{2} {r_1}^{2} + 3a^{4} z{r_1} + 3a^{5}}{4z^{5} {r_1}^{2}} - \frac{\left(4az^{4} {r_1}^{4} + 6a^{2} z^{3} {r_1}^{3} + 7a^{3} z^{2} {r_1}^{2} + 6a^{4} z{r_1} + 3a^{5}\right) \mathrm{e}^{-\frac{2z{r_1}}{a}}}{4z^{5} {r_1}^{2}}\right) d{r_1}
\end{aligned}
\end{equation}

Finally, after putting the bounds of integration from $r_1 = 0$ to $r_1 = \infty$, we obtain, 
\begin{equation}
    E' \approx (\frac{Z^3}{\pi a^3})^2 \frac{\pi }{3} G (\frac{h}{\lambda c}) \rho_0*(4\pi)^2\frac{15 \lambda ^{8}}{32z^{8}}
\end{equation}

The Python code used for this study is publicly available on GitHub at the following link: \url{https://github.com/rskhatiwada/perturbations/}.

\end{widetext}
\end{document}